\pdfoutput=1
\documentclass[prl,twocolumn,showpacs,preprintnumbers,amsmath,amssymb,
superscriptaddress,nofootinbib]{revtex4-1}
\usepackage{epsfig}
\usepackage{graphicx}
\usepackage{mathdots}
\usepackage{diagbox}

\newcommand{\be}{\begin{equation}}
\newcommand{\ee}{\end{equation}}
\newcommand{\bal}{\begin{aligned}}
\newcommand{\eal}{\end{aligned}}

\def\({\left(}
\def\){\right)}
\def\os{\overset}
\def\ap{\alpha}
\def\De{\Delta}

\begin{document}

\title{Simplex-like Structures of Maximally Supersymmetric Scattering Amplitudes}
\author{Junjie Rao}\affiliation{Max Planck Institute for Gravitational Physics (Albert Einstein Institute),
14476 Potsdam, Germany}
\date{\today}

\begin{abstract}
We elaborate the two-fold simplex-like structures of tree amplitudes in planar maximally
supersymmetric Yang-Mills ($\mathcal{N}\!=\!4$ SYM), through its connection to a mathematical structure
known as the positive Grassmannian.
Exploiting the reduced Grassmannian geometry and the matrix form of on-shell recursion relation in terms of super
momentum twistors, we manifest that tree amplitudes can be remarkably refined via the essential building blocks named
as fully-spanning cells. For a fixed number of negative helicities, an amplitude can be entirely captured
by finite, compact information of the relevant fully-spanning cells up to an arbitrarily large number of
external particles.
\end{abstract}

\maketitle

\section{Introduction}
\vspace{-10pt}

In recent years, enormous progress on scattering amplitudes has been made using various modern approaches beyond Feynman
diagrams (see e.g. \cite{Dixon:1996wi,Cachazo:2005ga,Henn:2014,Elvang:2015} for reviews).
In particular, amplitudes of $\mathcal{N}\!=\!4$ SYM in the planar limit are most understood due to its
unmatched symmetries. At both tree and loop levels, dual superconformal invariance manifested by
(super) momentum twistors \cite{Hodges:2009hk}, greatly facilitates the calculation of amplitudes and
loop integrands in planar $\mathcal{N}\!=\!4$ SYM \cite{ArkaniHamed:2010kv}.
Explicitly, this is realized by the momentum twistor version of BCFW
recursion relation \cite{Britto:2004ap,Britto:2005fq}, which constructs amplitudes solely from on-shell sub-amplitudes,
eliminating gauge redundancy as well as unphysical internal particles.

In the meanwhile, another unanticipated magic, namely the positive Grassmannian together with on-shell diagrams and
decorated permutations \cite{ArkaniHamed:2009dn,ArkaniHamed:2009vw,ArkaniHamed:2016},
provides new insights into the on-shell construction of amplitudes. This is mostly achieved
in the space of massless spinors, while transforming its entire machinery into momentum twistor space brings extra
complexity \cite{Bai:2014cna}, since each momentum twistor is not characterized by the momentum
of its literally corresponding particle, but a kinematic mixture of numerous adjacent particles.
It is this entanglement that trivializes momentum conservation,
so that we may concentrate on the pure kinematics separated from that universal constraint.
However, for non-planar $\mathcal{N}\!=\!4$ SYM, momentum twistors cannot be defined, while
on-shell diagrams still work \cite{Arkani-Hamed:2014bca,Franco:2015rma,Chen:2015bnt,Bourjaily:2016mnp},
in fact, the broad applicability of on-shell diagrams is independent of the number
of supersymmetries or spacetime dimensions \cite{Elvang:2014fja,Benincasa:2016awv}.

Back to planar $\mathcal{N}\!=\!4$ SYM,
to enhance the advantage brought by positive Grassmannian, we introduce another interesting excursion which brings even
more insights and richer structures of amplitudes \cite{Rao:2016out}, at tree level for the moment. It is a purely geometric
approach working in momentum twistor space without referring to on-shell diagrams and decorated permutations,
through establishing the exact correspondence
between Grassmannian geometric configurations and Yangian invariants generated by recursion.
The momentum twistor BCFW recursion relation is now presented in the matrix form of positive Grassmannian,
which can be nicely deduced from
positivity plus a minimal knowledge of momentum twistors. It is a simple linear algebra exercise to read off the
geometric configuration from the matrix representative of each BCFW cell, which can be
mapped back to its corresponding Yangian invariant directly. To encode this geometric information more compactly,
we need to introduce the reduced Grassmannian geometry for distinguishing linear dependencies of different ranks.
In addition, for BCFW cells it is also convenient to denote columns that are set to zero as ``empty slots'',
from which we will later reveal the two-fold simplex-like structures of tree amplitudes, as elaborated
in this letter. Then, for a fixed number of negative helicities, an amplitude can be entirely captured
by finite characteristic objects called fully-spanning cells
up to an arbitrarily large number of external particles.

\vspace{-8pt}
\section{Positive matrix form of momentum twistor BCFW recursion relation}
\vspace{-10pt}

\begin{figure}[h]
\includegraphics[scale=0.45]{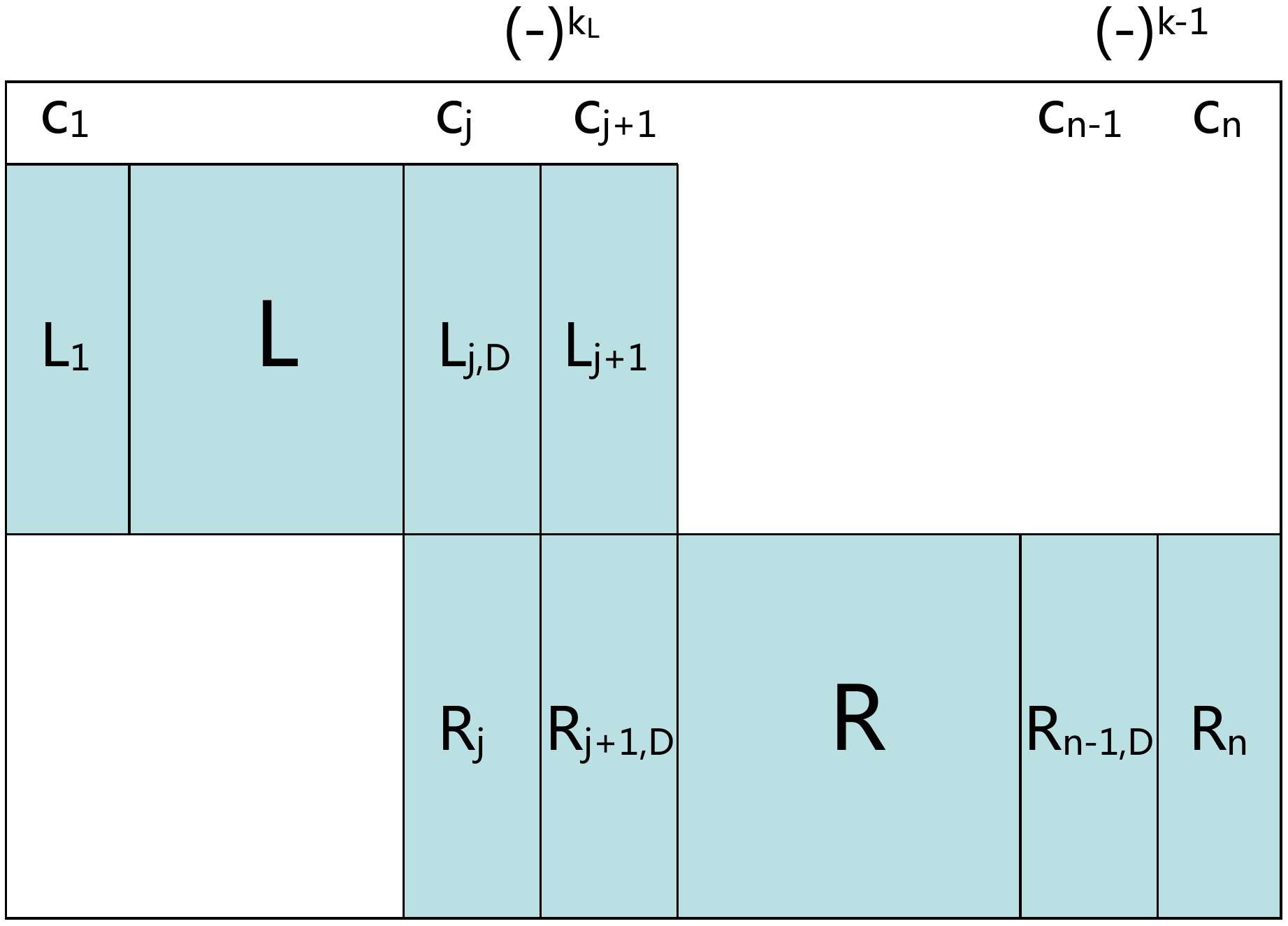}
\caption{Positive matrix form of momentum twistor BCFW recursion relation.
Sign factors $(-)^{k_\textrm{L}}$ and $(-)^{k-1}$ are associated to
the $c_j,c_{j+1}$ and $c_{n-1},c_n$ pairs respectively.
All the blank regions are filled with zero entries implicitly.} \label{fig-2}
\vspace{-12pt}
\end{figure}

To solely work in momentum twistor space, for a tree amplitude one can always first factor out the
maximally-helicity-violating (MHV) part, and the rest is the desired Yangian invariant we would like to address.
The BCFW recursion relation starts with the simplest Yangian invariants known as the 5-brackets,
then constructs more complex ones repeatedly with certain deformations that impose the on-shell condition of
internal particles. From the Grassmannian perspective, this can be deduced from positivity as a consequence
of pure geometry. We present the resulting matrix configuration in Figure \ref{fig-2}.

Let us give some explanation. For a $(k\times n)$ matrix $C_{\ap a}$ to have physical significance, where $(k+2)$ and $n$
are the numbers of negative helicities and total external particles respectively, $C$ has to be positive
(all of its ordered minors are positive or zero) and it obeys the orthogonal constraint $C_{\ap a}Z_a\!=\!0$ where $Z_a$'s
denote $n$ momentum twistors as kinematical data. If we have two such matrices $C_\textrm{L}$ and $C_\textrm{R}$,
we can construct a larger one by sewing them in some physical way, which induces deformations of the relevant columns in
these two sub-matrices. To parameterize the deformations, we need an additional row on the top and minimally it has five
entries to fulfill the constraint $C_{\ap a}Z_a\!=\!0$. In the geometric sense, the physical way above is nothing
but imposing positivity of this larger matrix! Explicitly, for $C_\textrm{L}$ spanning from column 1 to $(j\!+\!1)$ and
$C_\textrm{R}$ from $j$ to $n$ (see Fig. \ref{fig-2}), the deformed sub-columns with subscript `D' are given by

\vspace{-15pt}
\be
\bal
\!\!\!\!L_{j,\,\textrm{D}}&\!=\!L_j\!+\!\frac{c_j}{c_{j+1}}L_{j+1}, \\
\!\!\!\!R_{j+1,\,\textrm{D}}&\!=\!R_{j+1}\!+\!\frac{c_{j+1}}{c_j}R_j,~
R_{n-1,\,\textrm{D}}\!=\!R_{n-1}\!+\!\frac{c_{n-1}}{c_n}R_n,
\eal
\ee
\vspace{-12pt} \\
where the $c$'s are entries of the top row. To ensure positivity, extra sign factors $(-)^{k_\textrm{L}}$ and
$(-)^{k-1}$ must be associated to the $c_j,c_{j+1}$ and $c_{n-1},c_n$ pairs respectively.
After all $c$'s find their solutions in $C_{\ap a}Z_a\!=\!0$, the matrix above recovers
the BCFW product of Yangian invariants

\vspace{-15pt}
\be
[1\,j\,j\!+\!1\,n\!-\!1\,n]\,Y_\textrm{L}(1,\ldots,j,I)\,Y_\textrm{R}(I,j\!+\!1,\ldots,n\!-\!1,\widehat{n})
\ee
\vspace{-15pt} \\
where $\mathcal{Z}_I\!=\!\widehat{\mathcal{Z}}_{j+1}=\widehat{\mathcal{Z}}_j\!=\!(\,j\,j\!+\!1)\cap(n\!-\!1\,n\,1)$ and
$\widehat{\mathcal{Z}}_n\!=\!(n\!-\!1\,n)\cap(1\,j\,j\!+\!1)$.

Denoting the matrix in Fig. \ref{fig-2} as $Y_{n-1,\,j}$, we can express a general tree amplitude
(or Yangian invariant, precisely) as

\vspace{-25pt}
\be
Y^k_n=\sum_{i=k+3}^{n-1}\sum_{j=2}^{i-2}\,Y_{\,i,\,j}\,.
\ee
\vspace{-10pt} \\
Each matrix consists of a subset of BCFW cells of various $k_\textrm{L}$ and $k_\textrm{R}$ satisfying
$k_\textrm{L}\!+\!k_\textrm{R}\!=\!k\!-\!1\!\geq\!0$ and $0\!\leq\!k_\textrm{L,\,R}\!\leq\!n_\textrm{L,\,R}\!-\!4$,
with $k_\textrm{L}\!=\!0$ for $n_\textrm{L}\!=\!3$ as the only special case.
The ``sum'' of BCFW cells, or Grassmannian geometric configurations, in fact needs to be specified for avoiding
ambiguity of relative signs \cite{Bourjaily:2012gy,Olson:2014pfa}.
Such a discussion is presented in \cite{Rao:2016out}, where we
used some linear algebra trick to map BCFW cells back to Yangian invariants, and we plan to give a more
systemic treatment in the future. As we will soon see, this literal sum is indirectly justified by the cyclicity
of amplitudes via homological identities.

\vspace{-8pt}
\section{Reduced Grassmannian geometry}
\vspace{-10pt}

The matrix recursion relation generates more intricate geometric configurations beyond trivial single rows made of
five non-zero entries. For example, under the default recursion scheme, the N$^2$MHV $n\!=\!7$ amplitude is given by
(geometrically this is called a coutour)

\vspace{-12pt}
\be
Y^2_7\!=\![7]\!+\![5]\!+\![2]\!+\!(23)(45)\!+\!(23)(67)\!+\!(45)(71), \label{eq-1}
\ee
\vspace{-12pt} \\
where for instance, $[7]$ is a top cell with the 7th column removed, while $(23)(45)$ denotes vanishing minors
$(23)\!=\!(45)\!=\!0$. These BCFW cells are of $4k\!=\!k(n\!-\!k)\!-\!2\!=\!8$ dimensions,
for which kinematic and geometric degrees of freedom are equal (modulo GL$(k)$ invariance
and vanishing constraints for the latter).

Representing cells in this way is named as the Grassmannian geometry,
and in particular, $[i]$ which denotes the $i$-th column is null, is called an {\it empty slot}. Note that,
these symbols only make sense when $k,n$ are specified. For $k\!\geq\!3$, we need the {\it reduced}
Grassmannian geometry. For example, one N$^3$MHV $n\!=\!9$ BCFW cell is

\vspace{-15pt}
\be
(4\,\os{6}{\os{|}{5}}\,7)\,(8\,\os{1}{\os{|}{9}}\,2)
\ee
\vspace{-17pt} \\
where the ``upstair'' parts denote that, columns 5,6 are proportional and so are columns 9,1,
while as usual $(457)$ and $(892)$ are $3\times3$ vanishing minors. In this way,
linear dependencies of different ranks are distinguished unambiguously so that reading off its dimension is transparent.

We may apply (reduced) Grassmannian geometry to describe the homological identities, which are vanishing relations
between a number of ``boundary'' cells generated by the relevant $(4k\!+\!1)$-dimensional cells.
An example is the N$^2$MHV $n\!=\!7$ identity

\vspace{-15pt}
\be
\bal
0&=\partial(12) \\
&=-[1]\!+\![2]\!-\!(12)(34)\!+\!(12)(45)\!-\!(12)(56)\!+\!(12)(67),
\eal
\ee
\vspace{-10pt} \\
note that, we have discarded boundary cells that fail to have kinematical supports of $C_{\ap a}Z_a\!=\!0$,
which in this case are $(712)_2$ (abbreviation of (71)(12), and so forth) and $(123)_2$. But still,
we abuse the term ``homological'' here, while the actual kinematics
also matters.\footnote{We thank Jake Bourjaily for pointing out this subtlety.}

This is the only type of identities of $k\!=\!2,n\!=\!7$ up to a cyclic shift, and it guarantees the cyclicity of
N$^2$MHV $n\!=\!7$ amplitude via

\vspace{-15pt}
\be
Y^2_7\!-\!Y^2_{7,+1}\!=-\partial(23)\!-\!\partial(56)\!-\!\partial(71),
\ee
\vspace{-15pt} \\
where $Y^2_{7,+1}$ is the cyclicly shifted (by $+1$) counterpart of $Y^2_7$ in \eqref{eq-1}. Remarkably,
the cyclicity of N$^2$MHV amplitudes up to any $n$ can be shown in a similar but certainly more complicated way.
To manifest it demands the two-fold simplex-like structures of tree amplitudes, which we will immediately
exhibit in detail.

\vspace{-8pt}
\section{Triangle-like dissection of general N$^k$MHV amplitudes}
\vspace{-10pt}

Performing the BCFW recursion relation in its matrix form and using the representation of
reduced Grassmannian geometry, the two-fold simplex-like structures of tree amplitudes naturally emerge,
after some simple observation and refinement. As an appetizer, a general NMHV amplitude in terms of 5-brackets
is written as

\vspace{-15pt}
\be
Y^1_n\!=\!\sum_{i=4}^{n-1}\sum_{j=2}^{i-2}\,[1\,j\,j\!+\!1\,i\,i\!+\!1],
\ee
\vspace{-10pt} \\
now in terms of empty slots, it becomes

\vspace{-15pt}
\be
\bal
&Y^1_n\!= \\
&\!\!\!\(\!\begin{array}{c|cccccc}
[23\ldots n\!-\!4] & {} & {} & {} & {} & \!\!1 \\
\vdots~ & {} & {} & {} & \!\!\iddots & \!\!\vdots \\
{[23]}~ & {} & {} & \!\!1 & \!\cdots & \!\![\ldots n\!-\!2] \\
{[\textbf{2}]}~~ & {} & \!\!1 & \!\![5] & \!\cdots & \!\![5\ldots n\!-\!2] \\
1~~ & 1 & \!\![\textbf{4}] & \!\![45] & \!\cdots & \!\![45\ldots n\!-\!2] \\
\hline
{} & [\textbf{6}7\ldots n] & \!\![7\ldots n] & \!\![\ldots n] & \!\cdots & \!\!1
\end{array}\!\)\!, \label{eq-3}
\eal
\ee
\vspace{-12pt} \\
where $Y^1_n$ is the sum of all entries in the ``triangle'' above, and each entry is multiplied by its corresponding
vertical and horizontal factors, since we have maximally factored out common empty slots to manifest the pattern,
which is uniquely determined by the triple $(6,4,2)$ (in bold) for any $n$. This pattern will be later defined as
a {\it quadratic} growing mode.

The general N$^k$MHV amplitude directly follows a similar arrangement of the NMHV triangle, given by

\vspace{-15pt}
\be
\bal
&Y^k_n\!= \\
&\!\!\!\!\!\!\(\!\begin{array}{c|cccccc}
[2\ldots n\!-\!k\!-\!3] & {} & {} & {} & {} & \!\!1 \\
\vdots~ & {} & {} & {} & \!\!\iddots & \!\!\vdots \\
{[23]}~ & {} & {} & \!\!1 & \!\cdots & \!\!I_{n-2,3} \\
{[2]}~~ & {} & \!\!1 & \!\!I_{k+5,2} & \!\cdots & \!\!I_{n-1,2} \\
1~~ & 1 & \!\!I_{k+5,1} & \!\!I_{k+6,1} & \!\cdots & \!\!\!\!\!\!\!\!I_{n,1} \\
\hline
{} & [k\!+\!5\ldots n] & \!\![k\!+\!6\ldots n] & \!\![\ldots n] & \!\cdots & \!\!1
\end{array}\!\)\!, \label{eq-2}
\eal
\ee
\vspace{-12pt} \\
where $I_{i,1}$'s in the bottom row, each of which is a sum of BCFW cells, are the only essential objects to be identified,
since it is trivial to obtain $I_{i,1+j}$ by performing a \textit{partial} cyclic shift $i\!\to\!i\!+\!j$
except that label 1 is fixed, for all cells within $I_{i,1}$.
For example, from \eqref{eq-1} we already know $I_{7,1}$, then $I_{7,2}$ is simply given by

\vspace{-15pt}
\be
\bal
&I_{7,1}\!=\!
\(\!\begin{array}{cc}
\left\{\begin{array}{c}
\!\!(45)(71) \\
\!\!{[5]}~~~~
\end{array} \right. &
\!\!(23)\left\{\begin{array}{c}
\!\!(67) \\
\!\!(45)~~~~~~
\end{array} \right.
\end{array}\!\!\!\!\!\!\!\!\!\) \\
\to\,&I_{7,2}\!=\!
\(\!\begin{array}{cc}
\left\{\begin{array}{c}
\!\!(56)(81) \\
\!\!{[6]}~~~~
\end{array} \right. &
\!\!(34)\left\{\begin{array}{c}
\!\!(78) \\
\!\!(56)~~~~~~
\end{array} \right.
\end{array}\!\!\!\!\!\!\!\!\!\),
\eal
\ee
\vspace{-10pt} \\
where $I_{7,1}$ packs up four cells and so does $I_{7,2}$, note the common vanishing minors have been also factored out.

This is the triangle-like dissection of tree amplitudes, which isolates $I_{i,1}$ for further dissection.
Before that, let us digress to discuss how it refines the counting of BCFW cells. It is known that
the number of BCFW terms in tree amplitudes is given by

\vspace{-15pt}
\be
N^k_n\!=\!\frac{1}{n\!-\!3}\binom{n\!-\!3}{k}\binom{n\!-\!3}{k\!+\!1}.
\ee
\vspace{-10pt} \\
The double slicing (vertical and horizontal) in \eqref{eq-2} gives its {\it second order difference} as

\vspace{-15pt}
\be
\De^2 N^k_n\!=\!\De N^k_n\!-\!\De N^k_{n-1}\!=\!N^k_n\!-\!2N^k_{n-1}\!+\!N^k_{n-2},
\ee
\vspace{-15pt} \\
which is exactly the number of BCFW cells in $I_{n,1}$. For the first nontrivial case $k\!=\!2$, we have

\vspace{-15pt}
\be
\De^2 N^2_n\!=\!(n\!-\!5)^2, \label{eq-5}
\ee
\vspace{-15pt} \\
this will be useful as a highly nontrivial consistency check of the further simplex-like structure.

\vspace{-8pt}
\section{N$^2$MHV fully-spanning cells and solid simplices}
\vspace{-10pt}

Further dissecting $I_{i,1}$ reveals the following pattern:

\vspace{-15pt}
\be
\bal
I_{i,1}=\,&(\textrm{cells descend from }I_{i-1,1}) \\
&+(\textrm{new fully-spanning cells for }n\!=\!i),
\eal
\ee
\vspace{-10pt} \\
the first part of cells above follow simple patterns of the {\it solid simplices}, and so do the second
when $i$ is increased by one or more. The {\it fully-spanning cells} are named such that none of their $i$ columns
are removed when they first show up in $I_{i,1}$. Each time we increase $i$ by one, they are the only objects
need to be identified together with their {\it growing parameters}, which will uniquely determine their
``growing'' patterns in $I_{n,1}$ up to any $n$. Remarkably, the recursive growth of new fully-spanning cells terminates
at $n\!=\!4k\!+\!1$, as we will later see.

Explicitly, let us illustrate this pattern of $I_{i,1}$ for the family of N$^2$MHV amplitudes,
their fully-spanning cells are given by

\vspace{-17pt}
\be
\bal
G_{7,0}&=\left\{\begin{array}{c}
\!\!(45)(71) \\
\!\!{[5]}~~~~
\end{array} \right.~~~~~~~~~~~~~~~~\,(5)\,~~~~ \\
G_{7,1}&=(23)\left\{\begin{array}{c}
\!\!(67) \\
\!\!(45)~~~~~~
\end{array} \right.~~~~~~~~~~(6,4)~~~~~ \\
G_{8,1}&=\left\{\begin{array}{c}
\!\!(234)_2(678)_2~~~~~~~~~~~~(7,4)~~~~~ \\
\!\!(456)_2(781)_2~~~~~~~~~~~~(7,5)~~~~~ \\
\!\!(23)(456)_2(81)~~~~~~~~~\,(6,4)~~~~~ \\
\end{array} \right. \\
G_{9,2}&=\left\{\begin{array}{c}
\!\!(2345)_2(6789)_2~~~~ \\
\!\!(23)(4567)_2(891)_2
\end{array} \right.~~~\,(8,6,4) \label{eq-4}
\eal
\ee
\vspace{-10pt} \\
where $G_{i,m}$ is the part purely made of fully-spanning cells in $I_{i,1}$ and $m$ is its corresponding
{\it growing mode}, followed by their growing parameters (some cells share the same parameters). Note that
$[5]$ in $G_{7,0}$ above actually originates from $I_{6,1}$ as a top cell,
but for convenience it is put together with $(45)(71)$ as they share one parameter.

The meaning of growing modes and parameters can be seen from, for instance, how three sample cells below of constant,
linear and quadratic modes mutate as $i$ of $I_{i,1}$ increases, according to

\vspace{-15pt}
\be
(45)(71)\to[5](46)(81)\to[56](47)(91),
\ee
\vspace{-20pt}
\be
\bal
(23)(67)\to\,&[6](23)(78)\to[67](23)(89) \\
+&[4](23)(78)~\,+\![47](23)(89) \\
&~~~~~~~~~~~~~~~\,+\![45](23)(89),
\eal
\ee
\vspace{-12pt} \\
for $I_{7,1}$, $I_{8,1}$ and $I_{9,1}$, and

\vspace{-15pt}
\be
\bal
\!\!\!\!\!\!\!(2345)_2(6789)_2\!\to&[8](2345)_2(679\,10)_2\!\to\![89](2345)_2(67\,10\,11)_2 \\
+&[6](2345)_2(789\,10)_2\,+\![69](2345)_2(78\,10\,11)_2 \\
+&[4](2356)_2(789\,10)_2\,+\![67](2345)_2(89\,10\,11)_2 \\
&~~~~~~~~~~~~~~~~~~~~~~~~~+\![49](2356)_2(78\,10\,11)_2 \\
&~~~~~~~~~~~~~~~~~~~~~~~~~+\![47](2356)_2(89\,10\,11)_2 \\
&~~~~~~~~~~~~~~~~~~~~~~~~~+\![45](2367)_2(89\,10\,11)_2\,,
\eal
\ee
\vspace{-10pt} \\
for $I_{9,1}$, $I_{10,1}$ and $I_{11,1}$. Note the increasing numbers of empty slots induce partial cyclic shifts for
the associated cells, while maintaining their cyclic topologies,
similar to that of obtaining $I_{i,1+j}$ from $I_{i,1}$ in \eqref{eq-2}. We can further extract the key mathematical
objects that best describe all such patterns, namely the solid simplices.

\begin{figure}[h]
\includegraphics[scale=0.4]{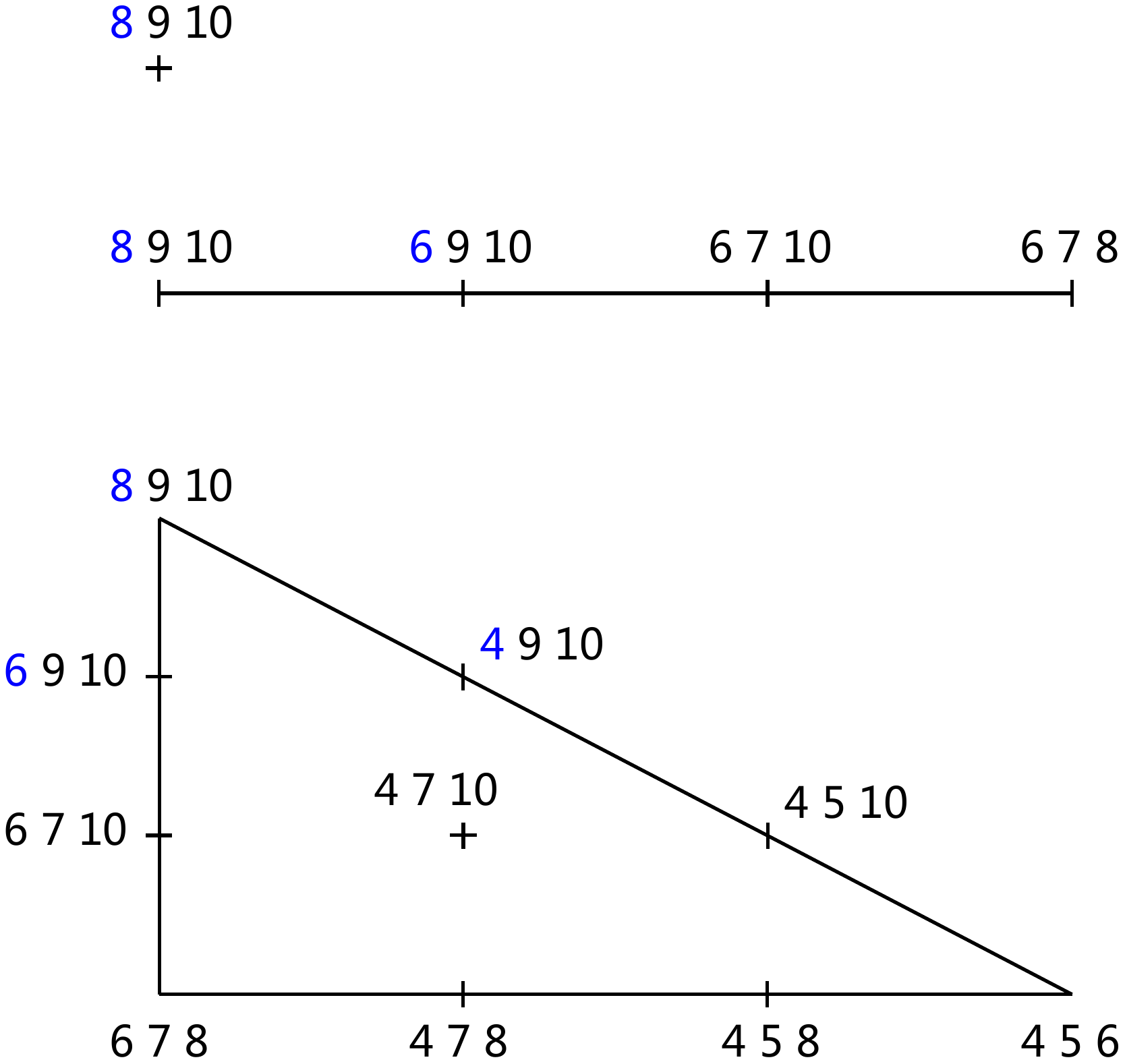}
\caption{Solid 0-, 1-, 2-simplices up to level 3, of growing \\
parameters $(8)$, $(8,6)$, $(8,6,4)$ respectively.} \label{fig-17}
\vspace{-5pt}
\end{figure}

A solid $m$-simplex is fully characterized by its growing mode $m$, $(m\!+\!1)$ growing parameters and {\it level}
which counts the empty slots at each point within it. The term ``solid'' means inside the simplex
there are also a number of points. In Figure \ref{fig-17}, we depict three solid simplices of constant, linear
and quadratic growing modes (0-, 1-, and 2-modes for short) up to level 3,
of growing parameters $(8)$, $(8,6)$, $(8,6,4)$ respectively.

Remarkably, the 2-mode with parameters $(6,4,2)$
exactly characterizes the NMHV triangle \eqref{eq-3}, which is extended to the general N$^k$MHV triangle-like dissection
in \eqref{eq-2}. Obviously, the two-fold simplex-like structures are closely related,
and the pattern of a 2-mode is manifest in the form which maximally factors out empty slots as \eqref{eq-3}.
General solid $m$-simplices similarly follow these patterns \cite{Rao:2016out}.
Now we can only concentrate on empty slots, while the induced geometric configurations can be trivially inferred
from their original cyclic topologies.

Let us immediately see the power of solid simplices for the N$^2$MHV case: back to \eqref{eq-4},
according to the growing modes and levels of fully-spanning cells, for any $n$, it is easy to count the terms
in $I_{n,1}$ as

\vspace{-15pt}
\be
2+2(n\!-\!6)+3(n\!-\!7)+2\cdot\frac{(n\!-\!7)(n\!-\!8)}{2}=(n\!-\!5)^2,
\ee
\vspace{-12pt} \\
which nicely matches $\De^2 N^2_n$ in \eqref{eq-5}.

\vspace{-8pt}
\section{Termination of the recursive growth of fully-spanning cells}
\vspace{-10pt}

Naturally, it is economical to only generate the fully-spanning cells along with their growing parameters,
for a given $k$. This is called the {\it refined} BCFW recursion relation \cite{Rao:2016out}, which constructs
fully-spanning cells solely from those of lower $k$'s, and it terminates at $n\!=\!4k\!+\!1$.
Explicitly, we find the numbers of fully-spanning cells for $k\!=\!1,2,3$ as summarized in the table below
(all the unspecified entries are zeros implicitly).

\vspace{4pt}
\begin{tabular}{|l|cccccccccc|}
  \hline
  \diagbox{~$k$}{$n$~~} & ~~5~ & ~6~ & ~7~ & ~8~ & ~9~ & ~10~ & ~11~ & ~12~ & ~13~ & ~14~~ \\
  \hline
  ~~1 & ~1 & {} & {} & {} & {} & {} & {} & {} & {} & {} \\
  \hline
  ~~2 & {} & 1 & 3 & 3 & 2 & {} & {} & {} & {} & {} \\
  \hline
  ~~3 & {} & {} & 1 & 7 & 18 & 27 & 26 & 15 & 5 & {} \\
  \hline
\end{tabular}
\vspace{6pt} \\
Note the first fully-spanning cell for any $k$ is a top cell, and it is the only one
in the anti-MHV sector ($n\!=\!k\!+\!4$).

\vspace{-8pt}
\section{Summary and Outlook}
\vspace{-10pt}

So far we have witnessed the concise profile of tree amplitudes in planar $\mathcal{N}\!=\!4$ SYM,
with the aid of the matrix form of BCFW recursion relation and reduced Grassmannian geometry.
The two-fold simplex-like structures are an extension following the same logic of,
e.g. \cite{Drummond:2008cr,ArkaniHamed:2009dg,Bourjaily:2010kw}. It is the simple Parke-Taylor formula \cite{Parke:1986}
of MHV tree amplitudes of gluons that first freed us from countless Feynman diagrams,
and up to this point, the similar idea has been extended to the solid simplices for general N$^k$MHV amplitudes
from the Grassmannian perspective, so that infinite terms now can be essentially captured by finite, compact information.

In the future, we will present how this formalism helps manifest the cyclicity of general N$^k$MHV amplitudes.
The NMHV sector has been solved in \cite{Rao:2016out}, so the first nontrivial case is the N$^2$MHV sector.
Also, we would like to explore how to extend it to, say, the 1-loop integrand level,
which is expected to be much more intricate.


\newpage
\begin{thebibliography}{99}


\bibitem{Dixon:1996wi}
L.~J.~Dixon,
\newblock (1996), arXiv:hep-ph/9601359.


\bibitem{Cachazo:2005ga}
F.~Cachazo and P.~Svrcek,
\newblock PoS {\bf RTN2005}, 004 (2005), arXiv:hep-th/0504194.


\bibitem{Henn:2014}
J.~M.~Henn and J.~C.~Plefka,
\newblock {\it Scattering Amplitudes in Gauge Theories} (Springer, 2014).


\bibitem{Elvang:2015}
H.~Elvang and Y.-t.~Huang,
\newblock {\it Scattering Amplitudes in Gauge Theory and Gravity} (Cambridge, 2015).


\bibitem{Hodges:2009hk}
A.~Hodges,
\newblock JHEP {\bf 1305}, 135 (2013), arXiv:0905.1473.


\bibitem{ArkaniHamed:2010kv}
N.~Arkani-Hamed, J.~L.~Bourjaily, F.~Cachazo, S.~Caron-Huot and J.~Trnka,
\newblock JHEP {\bf 1101}, 041 (2011), arXiv:1008.2958.


\bibitem{Britto:2004ap}
R.~Britto, F.~Cachazo and B.~Feng,
\newblock Nucl.\ Phys.\ B {\bf 715}, 499 (2005), arXiv:hep-th/0412308.


\bibitem{Britto:2005fq}
R.~Britto, F.~Cachazo, B.~Feng and E.~Witten,
\newblock Phys.\ Rev.\ Lett.\ {\bf 94}, 181602 (2005), arXiv:hep-th/0501052.


\bibitem{ArkaniHamed:2009dn}
N.~Arkani-Hamed, F.~Cachazo, C.~Cheung, and J.~Kaplan,
\newblock JHEP {\bf 1003}, 020 (2010), arXiv:0907.5418.


\bibitem{ArkaniHamed:2009vw}
N.~Arkani-Hamed, F.~Cachazo and C.~Cheung,
\newblock JHEP {\bf 1003}, 036 (2010), arXiv:0909.0483.


\bibitem{ArkaniHamed:2016}
N.~Arkani-Hamed, J.~L.~Bourjaily, F.~Cachazo, A.~B.~Goncharov, A.~Postnikov and J.~Trnka,
\newblock {\it Grassmannian Geometry of Scattering Amplitudes} (Cambridge, 2016).


\bibitem{Bai:2014cna}
Y.~Bai and S.~He,
\newblock JHEP {\bf 1502}, 065 (2015), arXiv:1408.2459.


\bibitem{Arkani-Hamed:2014bca}
N.~Arkani-Hamed, J.~L.~Bourjaily, F.~Cachazo, A.~Postnikov and J.~Trnka,
\newblock JHEP {\bf 1506}, 179 (2015), arXiv:1412.8475.


\bibitem{Franco:2015rma}
S.~Franco, D.~Galloni, B.~Penante and C.~Wen,
\newblock JHEP {\bf 1506}, 199 (2015), arXiv:1502.02034.


\bibitem{Chen:2015bnt}
B.~Chen, G.~Chen, Y.~K.~E.~Cheung, R.~Xie and Y.~Xin,
\newblock (2015), arXiv:1507.03214.


\bibitem{Bourjaily:2016mnp}
J.~L.~Bourjaily, S.~Franco, D.~Galloni and C.~Wen,
\newblock JHEP {\bf 1610}, 003 (2016), arXiv:1607.01781.


\bibitem{Elvang:2014fja}
H.~Elvang, Y.~t.~Huang, C.~Keeler, T.~Lam, T.~M.~Olson, S.~B.~Roland and D.~E.~Speyer,
\newblock JHEP {\bf 1412}, 181 (2014), arXiv:1410.0621.


\bibitem{Benincasa:2016awv}
P.~Benincasa and D.~Gordo,
\newblock (2016), arXiv:1609.01923.


\bibitem{Rao:2016out}
J.~Rao,
\newblock (2016), arXiv:1609.08627.


\bibitem{Bourjaily:2012gy}
J.~L.~Bourjaily,
\newblock (2012), arXiv:1212.6974.


\bibitem{Olson:2014pfa}
T.~M.~Olson,
\newblock JHEP {\bf 1508}, 120 (2015), arXiv:1411.6363.


\bibitem{Drummond:2008cr}
J.~M.~Drummond and J.~M.~Henn,
\newblock JHEP {\bf 0904}, 018 (2009), arXiv:0808.2475.


\bibitem{ArkaniHamed:2009dg}
N.~Arkani-Hamed, J.~Bourjaily, F.~Cachazo and J.~Trnka,
\newblock JHEP {\bf 1101}, 049 (2011), arXiv:0912.4912.


\bibitem{Bourjaily:2010kw}
J.~L.~Bourjaily, J.~Trnka, A.~Volovich and C.~Wen,
\newblock JHEP {\bf 1101}, 038 (2011), arXiv:1006.1899.


\bibitem{Parke:1986}
S.~J.~Parke and T.~R.~Taylor,
\newblock Phys.\ Rev.\ Lett.\ {\bf 56}, (1986) 2459.


\end{thebibliography}
\end{document}